\documentclass[a4paper,fleqn,usenatbib]{mnras}
\usepackage[T1]{fontenc}
\usepackage{ae,aecompl}
\usepackage{graphicx}	
\usepackage{amsmath}	
\usepackage{amssymb}	
\usepackage{ulem}

\title{Are There Pre-Main-Sequence/Black Hole X-ray Binaries?}
\author[Xu \& Li]{
Xiao-Tian Xu$^{1,2}$, 
Xiang-Dong Li$^{1,2}$\thanks{E-mail: lixd@nju.edu.cn}
\\
$^{1}$Department of Astronomy, Nanjing University, Nanjing 210046, China\\
$^{2}$Key Laboratory of Modern Astronomy and Astrophysics (Nanjing University), Ministry of Education, Nanjing 210046, China
}

\date{Accepted XXX. Received YYY; in original form ZZZ}

\pubyear{}
\begin{document}
\label{firstpage}
\pagerange{\pageref{firstpage}--\pageref{lastpage}}
\maketitle

\begin{abstract}
A large fraction of black hole low-mass X-ray binaries (BHLMXBs) are in short period orbits, which require strong orbital angular momentum loss during the previous evolutionary stage. \citet{Iva2006} put forward the possibility that some of the BHLMXBs may harbour pre-main-sequence (MS) donor stars, in order to explain the Li-overabundances in three BHLMXBs. In this work we investigate the evolution of low-mass pre-MS stars in binaries with a BH companion. We calculate the evolution of the spin and orbital periods, the stellar radius and the Roche-lobe radius during the pre-MS stage with the stellar evolution code MESA. We find that, because of the relatively slow rotation of the pre-MS star after the common envelope evolution and the long turnover time in the pre-MS stars, tidal torque is not always able to synchronize the pre-MS star, so the spin periods are generally longer than the orbital periods. Mass transfer can occur only for stars with mass larger than $\sim 1.2~M_\odot$, which experience expansion due to nuclear reaction after the Hayashi contraction phase. The effective temperatures and orbital periods of these systems  do not match the observations of BHLMXBs.
Our results show that the observed BHLMXBs with lithium overabundances are unlikely to host pre-main sequence donor stars.
\end{abstract}

\begin{keywords}
convection -- stars: abundances -- stars: black holes -- stars: pre-main sequence -- X-ray: binaries
\end{keywords}

\section{Introduction} \label{sec:intro}

There are 19 Galactic black hole (BH) X-ray binaries that are dynamically confirmed \citep{Rem2006,McC2006,Cas2014}. The majority of them are low-mass X-ray binaries (LMXBs), in which the mass of the companion star is $\lesssim 1\,M_\odot$. More than half of the BHLMXBs are characterized by relatively short orbital periods ($P_{\rm orb}\lesssim 1$ day). This indicates that the binary systems must have experienced the common envelope (CE) evolution \citep{Pac1976}, in which the low-mass companion star spirals into the envelope of the primary star (the progenitor of the BH), because of dynamically unstable mass transfer. After the CE phase the binary orbit is greatly reduced because the orbital energy is used to eject the envelope. If the binary can survive the supernova that produces the BH, it will finally evolve to be a LMXB \citep{vdH1983,Bha1991,Tau2006}.

A major difficulty of the standard picture of the formation of BHLMXBs is that, the CE evolution very likely leads to merger of the binary components, because the kinetic energy of the low-mass companion star is not enough to dissipate the envelope of the primary star in most cases \citep{Pod2003,Wang2016a,Wang2016b}. Several alternative scenarios have proposed in the literature \citep[see][for a review]{Li2015}. Detection of the CNO-processed material in XTE J1118+480 \citep{Has2002} suggests that the low-mass companion star could be originally an intermediate-mass star, which is more likely to survive the CE phase. This implies that some LMXBs may have evolved from intermediate-mass X-ray binaries \citep{Jus2006,Chen2006}. Meanwhile, overabundances of lithium were reported in three BHLMXBs \citep{Mar1994,Mar1996}. \citet{Mac2005} proposed that the overabundance can be explained by the tidally locked rotation of the secondary stars, which lead to slower lithium destruction rates, while \citet{Iva2006} explained the  abundance of lithium by assuming that the secondary star is a pre-main sequence (MS) star. In the latter case, as the star contracts toward the MS, the strong magnetic field brakes the star's rotation, which can dissipate the orbital angular momentum through tidal interaction and drive the binary to contact.

Despite the possible criticisms of the pre-MS scenario as listed in \citet{Iva2006}, a large fraction of BHMXBs should have evolved through the stage when the companion star was a pre-MS star, since
the pre-MS lifetime of a $1\,M_{\odot}$ star is about $10^7$ yr, while the lifetime  of a $>25 M_{\odot}$ primary star is at most a few $10^6$ yr.
Thus it is interesting to see whether the binary can appear as a LMXB. If the secondary star is tidally locked, magnetic braking can efficiently  shrink the orbit to trigger mass transfer.  However, in this work we demonstrate that it is not always the case, and whether mass transfer can occur critically depends on the mass and the evolutionary state of the pre-MS star.

The rest of the paper is organized as follows. In section 2, we describe theoretical analysis on the spin and orbital evolution. In section 3, we present numerically calculated results of the binary evolution. We summarize and discuss our work in section 4.

\section{Model}
\subsection{Spin evolution}
The rate of change of the spin angular momentum $\dot{J}_2$ of the pre-MS star is controlled by the following equation,
\begin{equation}
\dot{J}_2=\dot{I}_2\Omega_2+I_2\dot{\Omega}_2=\dot{J}_{\rm 2,MB}+\dot{J}_{\rm 2,tid},
\label{jdot2}
\end{equation}
with $\dot{X}=\partial X/\partial t$. Here $I_2$ and $\Omega_2$ are the momentum of inertia and the angular velocity of the star, $\dot{J}_{\rm 2,MB}$ and $\dot{J}_{\rm 2,tid}$ represent the angular momentum loss rate due to  magnetic braking and tidal torque, respectively. We assume that the star rotates rigidly.

\begin{figure}
\includegraphics[width=\columnwidth]{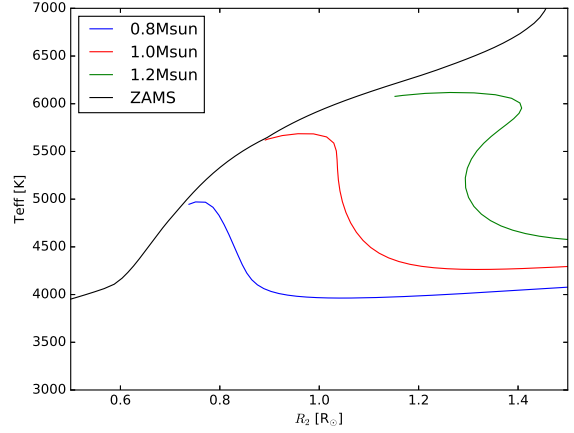}
\caption{
The evolutionary tracks of non-rotating pre-MS star in the $T_{\rm eff}-R_2$ plane. The black line represents the MS in the $T_{\rm eff}-R_2$ plane. The blue, red, and green line represents the evolutionary track with stellar mass $M_2=0.8\,M_{\odot}$, $1.0\,M_{\odot}$, and $1.2\,M_{\odot}$ respectively.
\label{f1}}
\end{figure}

Let's first examine the effect of stellar evolution. When evolving toward ZAMS, a pre-MS star spins up or down due to the change in its radius. In Fig.~\ref{f1}, we plot the evolutionary tracks of three non-rotating pre-MS stars in the effective temperature ($T_{\rm eff}$) - stellar radius ($R_2$) plane. The blue, red, and green lines represent the evolutionary tracks  with the stellar mass $M_2=0.8\,M_{\odot}$, $1.0\,M_{\odot}$, and $1.2\,M_{\odot}$ respectively, and the black line denotes the ZAMS. A large part of the pre-MS lifetime of the stars is spent on the Hayashi track, which is the horizontal part of the evolutionary tracks in Fig.~\ref{f1}. Note that the stars behave differently after leaving the Hayashi track. The radius of the $0.8\,M_{\odot}$ star continues decreasing until the star reaches ZAMS,
the radius of the $1.0\,M_{\odot}$ star stays roughly the same for some time before it decreases, and the radius of the
$1.2\,M_{\odot}$ star first increases then decreases.

Taking into account stellar rotation, we can write the hydrostatic equilibrium equation for the stellar structure to be \citep{Mae2009}
\begin{equation}
\frac{1}{\rho}\nabla P_{\rm p} = -\nabla (\Phi_{\rm g}-\Phi_{\rm \Omega}),
\label{hyd}
\end{equation}
where $\rho$ is the material density, $P_{\rm p}$ gas and radiation pressure, $\Phi_{\rm g}$ the gravitational potential, and $\Phi_{\rm \Omega}$ the centrifugal potential.  Obviously rotation can partly counteract the gravitational potential, leading to a larger radius.

\begin{figure}
\includegraphics[width=\columnwidth]{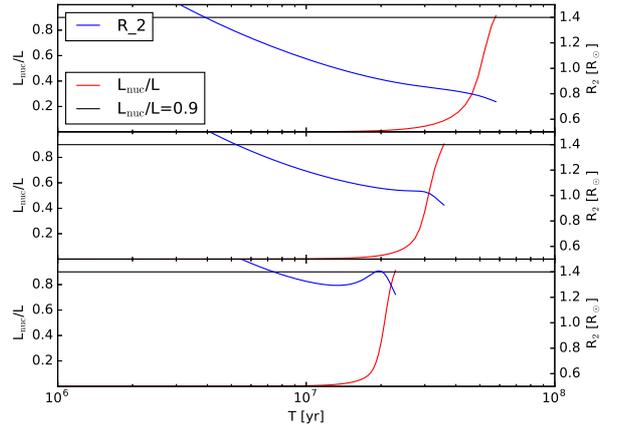}
\caption{
The evolutionary tracks of $L_{\rm nuc}/L$ and $R_2$ over evolutionary time $T$. the upper, middle, lower panel are calculated with $M_2=0.8\,M_{\odot}$, $1.0\,M_{\odot}$, and $1.2\,M_{\odot}$ respectively. The red solid lines represent the evolution of $L_{\rm nuc}/L$. The blue solid lines represent the evolution of $R_2$. The black solid line represents $L_{\rm nuc}/L=0.9$, marking the birth of a MS star.
\label{f2}}
\end{figure}

Near the ZAMS, $T_{\rm eff}$ increases significantly while the evolution of $R_2$ slows down.  Here we highlight an important feature of the track with $M_2=1.2~M_\odot$ that the star experiences an expansion phase after  shrinking on the Hayashi line (we call it the ``Z-shape" track). Ignition of nuclear reaction is responsible for this Z-shape feature. In Fig. \ref{f2}, we plot the evolution of the ratio of the nuclear luminosity and the total luminosity $L_{\rm nuc}/L$ and the stellar radius $R_2$ against the evolutionary time $T$. In this figure, the upper, middle and lower panels show the cases of $M_2=0.8\,M_{\odot}$, $1.0\,M_{\odot}$, and $1.2\,M_{\odot}$, respectively. The red and blue solid lines represent the evolutionary tracks of $L_{\rm nuc}/L$ and $R_2$ respectively, the black solid line denotes $L_{\rm nuc}/L=0.9$, marking the formation of a ZAMS star. In the case of $M_2=0.8\,M_{\odot}$ and $1.0\,M_{\odot}$, nuclear burning proceeds relatively gently, while $L_{\rm nuc}$ increases rapidly in the case of $M_2=1.2\,M_\odot$, which results in temporary expansion of the star.

The $\dot{J}_{\rm 2,MB}$ term in Eq.~(\ref{jdot2}) represents the effect of magnetic braking. Here we adopt the traditional view that magnetic braking is turned off if the star does not have a radiative core. To evaluate the angular momentum loss rate caused by stellar wind coupled with a magnetic field, we adopt the formula derived by \citet{Jus2006},
\begin{equation}\label{mb}
\dot{J}_{\rm 2,MB}=-\frac{2\pi}{P_{\rm spin}}B_{\rm s}
R_2^{13/4}\dot{M}_{\rm w}^{1/2}(GM_2)^{-1/4},
\end{equation}
where $P_{\rm spin}$, $B_{\rm s}$, and $\dot{M}_{\rm w}$ are respectively the spin period, the surface magnetic field strength, and the wind mass loss rate of the star, and $G$ is the gravitational constant.

A pre-MS star is characterized by strong wind and magnetic field, for which we follow the parameterizations suggested by \citet{Iva2006}. According to the dynamo mechanism \citep{Par1971,Hin1989,Meu1997}, the magnitude of $B_{\rm s}$ is related to the Rossby number $\tau_{\rm B}$ in the form $B_{\rm s}\propto \tau_{\rm B}^{-\alpha}$, where $\alpha$ is taken to be 1, and
\begin{equation}
\tau_{\rm  B}=\frac{P_{\rm spin}}{\tau_{\rm c}}. \end{equation} Here $\tau_{\rm c}$ is the turnover time of the convection zone in the star, expressed as, \begin{equation} \tau_{\rm c}=\frac{H_{\rm p}}{v_{\rm conv}}, \end{equation} where $H_{\rm p}$ is the pressure scale height and $v_{\rm conv}$ is the velocity of the convection cell.

Then, the relation between $B_{\rm s}$, $\tau_{\rm c}$ and $P_{\rm spin}$ can be written in the following form,
\begin{equation}
\frac{B_{\rm s}}{B_{\rm s,0}}=f_{\rm B}\left(\frac{\tau_{\rm c}}{800 {\rm ~day}}\right)\left(\frac{1{\rm ~day}}{P_{\rm spin}}\right),
\label{equ6}
\end{equation}
where $B_{\rm s,0}$ is the initial magnetic field, and $f_{\rm B}$ is a constant. In this work, we neglect the possible influence of the magnetic field on the stellar structure. Observations suggest that $B_{\rm s}$ of T Tauri stars  can reach up to $\sim 5$ kG  \citep{OSu2005,Sym2005}, so we take $B_{\rm s,0}=10$ kG
as in \citet{Iva2006}. As for the wind mass loss rate of pre-MS stars, \citet{Iva2006} argued that, although $B_{\rm s}$ and $\dot{M}_{\rm w}$ may not be physically related, $B_{\rm s}$ and $\dot{M}_{\rm w}$  may decay with a similar law. Thus, one can write \begin{equation}
\frac{\dot{M}_{\rm w}}{\dot{M}_{\rm w,0}}=f_{\rm w}\frac{B_{\rm s}}{B_{\rm s,0}},
\end{equation}
where $\dot{M}_{\rm w,0}$ is the initial stellar wind loss rate, and $f_{\rm w}$ is a constant. For the Sun, the mass loss rate is very low $\sim 10^{-13.6}~M_\odot\,{\rm yr}^{-1}$ \citep{Egg2006},  while the wind loss rate of a pre-MS star can be $\sim 10^{-9}-10^{-8}~M_\odot\,{\rm yr}^{-1}$ \citep{Lor2006}. Here we adopt $(\dot{M}_{\rm w,0},\,f_{\rm w})=(10^{-9}~M_\odot\,{\rm yr}^{-1},\,1)$ \citep{Iva2006}. In addition, we take into account the effect of rotation on  the stellar wind. Because of the centrifugal force, material on the surface of a rapidly rotating star has more chance to escape, so the wind loss rate is enhanced by a factor \citep{Heg2000},
\begin{equation}
\frac{\dot{M}_{\rm w}}{\dot{M}_{\rm w}(\Omega_2=0)}=
(1-\Omega_2/\Omega_{\rm crit})^{-n},
\end{equation}
where $n$ is a constant with typical value 0.43 and $\Omega_{\rm crit}$ is the critical angular velocity given by
\begin{equation}
\Omega_{\rm crit}^2=\left(1-\frac{\kappa L}{4\pi c GM_2}\right)\frac{GM_2}{R_2^3},
\end{equation}
where $c$ is the speed of light, and $\kappa$ is the opacity at the surface of the star.

\begin{figure}
\includegraphics[width=\columnwidth]{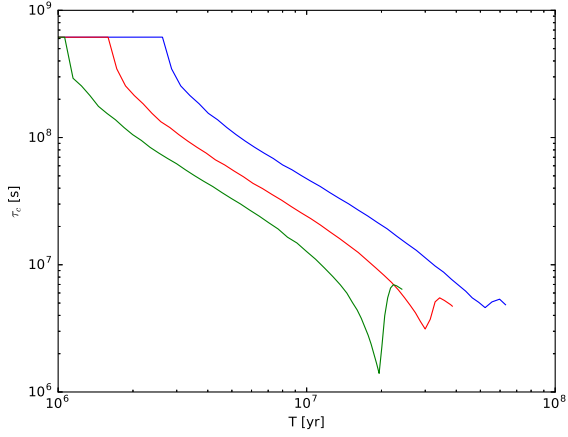}
\caption{
The evolution of $\tau_{\rm c}$ in the pre-MS state. The blue, red, and green line represents the evolutionary track with $M_2=0.8\,M_{\odot}$, $1.0\,M_{\odot}$, and $1.2\,M_{\odot}$ respectively.
\label{f3}}
\end{figure}

For the turnover time $\tau_{\rm c}$, we adopt the empirical formula given by \citet{Sad2017},
\begin{equation}
\begin{split}
{\rm lg}(\tau_{\rm c}/ 1{\rm ~s})&=8.79-2|{\rm lg}(m_{\rm cz})|^{0.349}-0.0194|{\rm lg}(m_{\rm cz})|^2\\
&-1.62{\rm min}[|{\rm lg}(m_{\rm cz})+8.55|,0],                                                                                                                             \end{split}                            \end{equation}
where $m_{\rm cz}=M_{\rm cz}/M_2$ and $M_{\rm cz}$ is the mass of the convection zone. In Fig. \ref{f3}, we plot the evolution of $\tau_{\rm c}$ for $0.8\,M_\odot$, $1.0\,M_\odot$, and $1.2\,M_\odot$ non-rotating pre-MS stars with the blue, red, and green lines, respectively. It can be seen that generally $\tau_{\rm c}$ is
much longer in the pre-MS stage than in the MS stage.

The turnover time $\tau_{\rm c}$ plays an important role in the evolution of a pre-MS/BH binary. It affects not only the evolution of $B_{\rm s}$ and $\dot{M}_{\rm w}$ but also the efficiency of tidal torque. The changing rate of the spin angular velocity of the pre-MS star induced by tidal torque is evaluated by
\begin{equation}
\frac{d\Omega_2}{dt}=\frac{\Omega_{\rm orb}-\Omega_2}{\tau_{\rm tid}},
\label{tid}
\end{equation}
where $\Omega_{\rm orb}$ is the orbital angular velocity, $\Omega_{\rm orb}=2\pi/P_{\rm orb}$, and $\tau_{\rm tid}$ is the tidal time-scale. We adopt the Eq. (11) in \citet{Hut1981} to evaluate $\tau_{\rm tid}$,
\begin{equation}
\tau_{\rm tid}=\left[3\frac{k}{T}\frac{q^{-2}}{r_{\rm g}^2}
\left(\frac{R_2}{a}\right)^6\right]^{-1},
\label{ttid}
\end{equation}
where $r_{\rm g}^2=I/(M_2R_2^2)$,  $q$ the mass ratio of the binary, $q=M_2/M_{\rm BH}$ ($M_{\rm BH}$ is the BH mass),  $a$  the binary separation, and $k/T$ evaluated by the following formula \citep{Hur2002},
\begin{equation}
\frac{k}{T}=\frac{2}{21}\frac{f_{\rm conv}}{\tau_{\rm c}}
\frac{M_{\rm cz}}{M_2}{\rm ~yr}^{-1},
\label{kt}
\end{equation}
and
\begin{equation}
f_{\rm conv}={\rm min}\left[1,\left(\frac{P_{\rm tid}}
{2\tau_{\rm c}}\right)^2\right],
\label{fconv}
\end{equation}
where
\begin{equation}
\frac{1}{P_{\rm tid}}=\left|\frac{1}{P_{\rm orb}}-
\frac{1}{P_{\rm spin}} \right|.
\label{equ15}
\end{equation}
For stars with a convective envelope, one can see that $\tau_{\rm tid}\propto\tau_{\rm c}f_{\rm conv}^{-1}$. Thus, tidal torque in a BH binary is weaker for a pre-MS secondary than for a MS secondary with the same mass. Also note that $f_{\rm conv}$ sensitively depends on $P_{\rm spin}$ and $P_{\rm orb}$: generally $f_{\rm conv}\ll 1$ before synchronization and $f_{\rm conv}=1$ after synchronization.

\subsection{Orbital evolution}
The orbital angular momentum $\dot{J}_{\rm  orb}$ is defined as
\begin{equation}
J_{\rm or b}=\frac{M_2M_{\rm BH}}{M}\sqrt{GMa},
\label{jorb}
\end{equation}
where $M=M_2+M_{\rm BH}$. Taking the logarithmic derivative of Eq.~(\ref{jorb}) gives
\begin{equation}
\frac{\dot{a}}{a}=2\frac{\dot{J}_{\rm orb}}{J_{\rm orb}}-2\frac{\dot{M}_2}{M_2}-
2\frac{\dot{M}_{\rm BH}}{M_{\rm BH}}+\frac{\dot{M}}{M}.
\label{adot}
\end{equation}
The above equation demonstrates that the orbital evolution is determined by mass and angular momentum loss and transfer. To maintain stable mass transfer, the evolution of $a$ has to meet two requirements: $R_2=R_{\rm L,2}$ (where $R_{\rm L,2}$ is the radius of the Roche Lobe of the donor star) and $\dot{R}_2=\dot{R}_{\rm L,2}$. In the first place, we consider the effect of mass loss and exchange. The total mass loss rate $\dot{M}_2$ of the pre-MS star consists of two parts: the stellar wind loss rate $\dot{M}_{\rm w}$ and the mass transfer rate $\dot{M}_{\rm 2,rlo}$ via Roche-lobe overflow, i.e., \begin{equation}
-\dot{M}_2=\dot{M}_{\rm w}+\dot{M}_{\rm 2,rlo}.
\end{equation}
Here the mass transfer rate is evaluated based on \citet{Rit1988}'s scheme, \begin{equation}
\dot{M}_{\rm 2, rlo}=\dot{M}_{\rm rlo,0}{\rm exp}\left[\frac{R_2-R_{\rm L,2}}{H_{\rm p}\gamma(q)}\right],
\label{mdot}
\end{equation}
where $\dot{M}_{\rm rlo,0}$ is given by
\begin{equation}
\dot{M}_{\rm rlo,0}=\frac{1}{e^{1/2}}\rho c_{\rm th}Q,
\end{equation}
and $\gamma(q)$ is a function of the mass ratio. In Eq.~(20), $Q$ is the cross section of the mass flow via the L1 point, and $\rho$ and $c_{\rm th}$ are the matter density and sound speed evaluated at the surface of the star, respectively. The Roche-lobe radius $R_{\rm L,2}$ is given by the formula proposed by \citet{Egg1983},
\begin{equation}
\frac{R_{\rm L,2}}{a}=\frac{0.49q^{2/3}}{0.6q^{2/3}+{\rm ln}(1+q^{1/3})}.
\end{equation}
We assume that the accretion rate of the BH is limited by the Eddington accretion rate $\dot{M}_{\rm Edd}$, that is, \begin{equation}
\dot{M}_{\rm BH} = {\rm min}[\dot{M}_{\rm 2,rlo}, \dot{M}_{\rm Edd}],
\end{equation}
where
\begin{equation}
\dot{M}_{\rm Edd}=2.6\times 10^{-7} {\rm ~M_\odot~yr^{-1}}
\left(\frac{M_{\rm BH}}{10{M_\odot}}\right)\left(\frac{\eta}{0.1}\right)^{-1}\left(\frac{1+X}{1.7}\right)^{-1}.
\end{equation}
Here $X$ is the mass fraction of hydrogen in the accreting gas and $\eta$ is the  radiation efficiency
\begin{equation}
\eta = 1-\sqrt{1-\left(\frac{M_{\rm BH}}{3M_{\rm BH}^0}\right)^{2}},
\end{equation}
where $M_{\rm BH}^0$ is the initial BH mass \citep{Pod2003}.

The change in the orbital angular momentum $\dot{J}_{\rm orb}$ is caused by gravitational radiation, mass loss, and tidal interaction, which are represented by the the three terms on the right-hand-side of the following equation, respectively
\begin{equation}
\dot{J}_{\rm orb}=\dot{J}_{\rm orb,GR}+\dot{J}_{\rm orb,ML}+\dot{J}_{\rm orb,tid},
\label{jorbdot}
\end{equation}  For angular momentum loss due to gravitational radiation, we adopt the standard formula given by \citet{Lan1975},
\begin{equation}
\dot{J}_{\rm orb,GR}=-\frac{32}{5c^5}\left(\frac{2\pi G}{P_{\rm orb}}\right)^{7/3}
\frac{(M_2M_{\rm BH})^2}{(M_{\rm BH}+M_2)^{2/3}}.
\end{equation} We assume that stellar wind matter carries the specific orbital angular momentum of the pre-MS star and the ejected matter in the case of super-Eddington accretion carries the specific angular momentum of the BH. Then, $\dot{J}_{\rm orb,ML}$  is given by
\begin{equation}
\dot{J}_{\rm orb,ML}=-\dot{M}_{\rm w}a^2{\Omega_{\rm orb}}\left(\frac{M_{\rm BH}}{M_2+M_{\rm BH}}\right)^2\ \ {\rm if}\ \dot{M}_{\rm 2,rlo}<\dot{M}_{\rm Edd},
\end{equation}
or
\begin{equation}
\begin{split}
\dot{J}_{\rm orb,ML}=-\left[\dot{M}_{\rm w}M_{\rm BH}^2
+(\dot{M}_{\rm 2,rlo}-\dot{M}_{\rm Edd})M_2^2\right]\frac{a^2\Omega_{\rm orb}}{(M_2+M_{\rm BH})^2}\\
\ \ {\rm if}\ \dot{M}_{\rm 2,rlo}>\dot{M}_{\rm Edd}.
\end{split}
\end{equation}
Finally, the orbital angular momentum is transferred to the pre-MS star by tidal torque, i.e.
\begin{equation}
\dot{J}_{\rm orb,tid}=-\dot{J}_{\rm 2,tid}.
\end{equation}
If the rotation of the secondary star is locked by tidal interaction and its spin angular momentum loss via magnetic braking is replenished by tidal torque immediately, magnetic braking is able to extract the orbital angular momentum efficiently. However, the tidal torque for a non-synchronized pre-MS star is weaker compared with that for a synchronized MS star, due to its relatively longer $\tau_{\rm c}$
or smaller $f_{\rm conv}$  
after the CE evolution (see section 4),  while magnetic braking becomes stronger due to stronger magnetic field and stellar wind. Therefore one needs to carefully check
whether tidal interaction can synchronize the rotation of the pre-MS star.

\section{Results}
In this section, we present the numerically calculated results of the binary evolution by use of the Modules for Experiments in Stellar Astrophysics (MESA; version number 10108) \citep{Pax2011,Pax2013,Pax2015,Pax2017}.

\subsection{Method and numerical setup}

Our calculation is divided into following steps. Firstly, we use the star module of MESA to generate a pre-MS star model with  a specific age ($T_0$) and mass ($M_2$),  assuming that the high-mass primary star evolves into a BH at $T_0$. Secondly, we use this stellar model as the input model for the binary module of MESA.
Since the focus of our work is whether some of the observed X-ray binaries currently host pre-MS stars, we assume that the pre-MS star fills its Roche-lobe at the birth of the BH, that is $R_2=R_{\rm L,2}$ at  $T=T_0$. If it is not the cases, then a fair amount of time will be needed for the orbit to shrink so that the star can contact its Roche-lobe.  Thus, although the binary may at some point reach Roche-lobe overflow, the star will likely be on the MS and the system will not be of the type that is addressed in this paper (see Fig.~12 below).

With this condition one can derive the orbital period at  $T_0$. Obviously, the longer the initial orbital period, the more time the pre-MS star has to reach the MS.

Our numerical setup is described as follows. The numerical timestep $\Delta T$ is set to be
\begin{equation}
\Delta T={\rm min}\left[10^{-3}|\frac{J_2}{\dot{J}_{\rm 2,MB}}|,10^{-3}|\frac{J_2}{\dot{J}_{\rm 2,tid}}|,
\Delta T_{\rm MESA}\right]
\end{equation}
where $\Delta T_{\rm MESA}$ is the default timestep used in MESA. We adopt the following input parameters:
\begin{itemize}
\item The initial mass of the BH: $M_{\rm BH,0}=7\,M_\odot$;
\item The initial age of the pre-MS star (or the birth time of the BH): $T_0=(0.5\times 10^7 {\rm ~yr},
1\times10^7 {\rm ~yr}, 1.5\times10^7 {\rm ~yr})$;
\item The initial mass of the pre-MS star: $M_2=(0.8\,{M_\odot}, 1.0\,{M_\odot}, 1.2\,{M_\odot})$;
\item The initial spin period of the pre-MS star: $P_{\rm spin}=1$ day;
\item The initial magnetic field and wind loss rate: $B_{\rm s,0}=10$ kG, $f_{\rm B}=(1, 3, 10)$; $\dot{M}_{\rm w,0}=-10^{-9}M_\odot$\,yr$^{-1}$, $f_{\rm w}=1$;
\item The initial orbital period $P_{\rm orb }$ is set by $R_2=R_{\rm L,2}$ at $T_0$.
\end{itemize}
We terminate our calculation when the star evolves into a MS star,  marked by $L_{\rm nuc}/L = 0.9$.

\subsection{Evolution of the spin and orbital periods}

\begin{figure}
\includegraphics[width=\columnwidth]{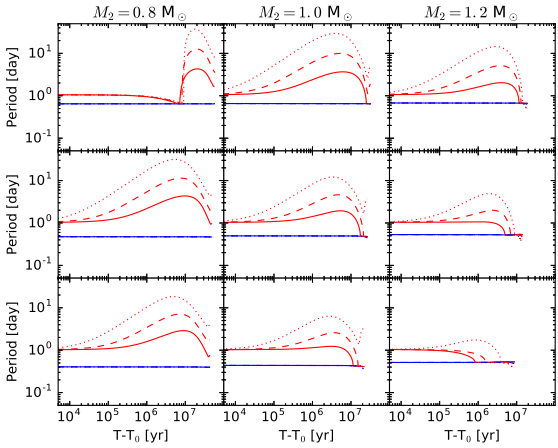}
\caption{
The evolution of $P_{\rm spin}$ (red lines) and $P_{\rm orb}$ (blue lines).
In the first, second, and third columns we take $M_2=0.8~M_\odot$, $1.0~M_\odot$, and $1.2~M_\odot$ respectively, and in the first, second, and third rows we take $T_0=0.5\times10^{7}$ yr, $1.0\times10^{7}$ yr, and $1.5\times10^{7}$ yr respectively. In each panel the solid, dashed, and dotted lines represent the evolution
with $f_{\rm B}=1$, 3, and 10, respectively.
\label{f4}}
\end{figure}

In Fig.~\ref{f4}, we plot the calculated evolution of $P_{\rm spin}$ and $P_{\rm orb}$ with the red and blue lines, respectively. In the first, second, and third columns we take $M_2=0.8~M_\odot$, $1.0~M_\odot$, and $1.2~M_\odot$ respectively, and in the first, second, and third rows we take $T_0=0.5\times10^{7}$ yr, $1.0\times10^{7}$ yr, and $1.5\times10^{7}$ yr respectively. In each panel the solid, dashed, and dotted lines represent the evolution with $f_{\rm B}=1$, 3, and 10, respectively. We can see that the spin periods change significantly during the evolution while the orbital periods keep nearly constant. Except the cases with $(M_2,\,T_0)=(0.8~M_{\odot},\,0.5\times10^7\,{\rm yr})$ and $(1.2~M_{\odot},\,1.5\times10^7\,{\rm yr})$,
there is a general trend where the pre-MS star has a spin-down phase followed by a spin-up phase.
In the spin-down phase, the evolution of $P_{\rm spin}$ is controlled by magnetic braking. The maximum $P_{\rm spin}$ is longer for less massive stars and larger values of $f_{\rm B}$. Then, with decreasing $(B_{\rm s},~\dot{M}_{\rm w},~\tau_{\rm c})$ and increasing $f_{\rm conv}$,  tidal torque starts to dominate the evolution. In the case of $(M_2,\,T_0)=(0.8~M_{\odot},\,0.5\times10^7\,{\rm yr})$, the evolution of $P_{\rm spin}$ firstly experiences a spin-up because magnetic braking is assumed to be turned off due to the absence of a radiative core. In the case of $(M_2,\,T_0)=(1.2~M_{\odot},\,1.5\times10^7\,{\rm yr})$, tidal torque dominates the spin evolution initially for $f_{\rm B}=$1 and 3, because the relatively weaker magnetic field of the pre-MS star according to Eq.~(\ref{equ6}) and Fig.~\ref{f3}. One can also see that, when approaching the MS, stars with strong magnetic fields spin down again because of the increase in $\tau_{\rm c}$. In the case of weak magnetic field,  tidal torque is more likely to finally synchronize the rotation of the pre-MS star.

\begin{figure}
\includegraphics[width=\columnwidth]{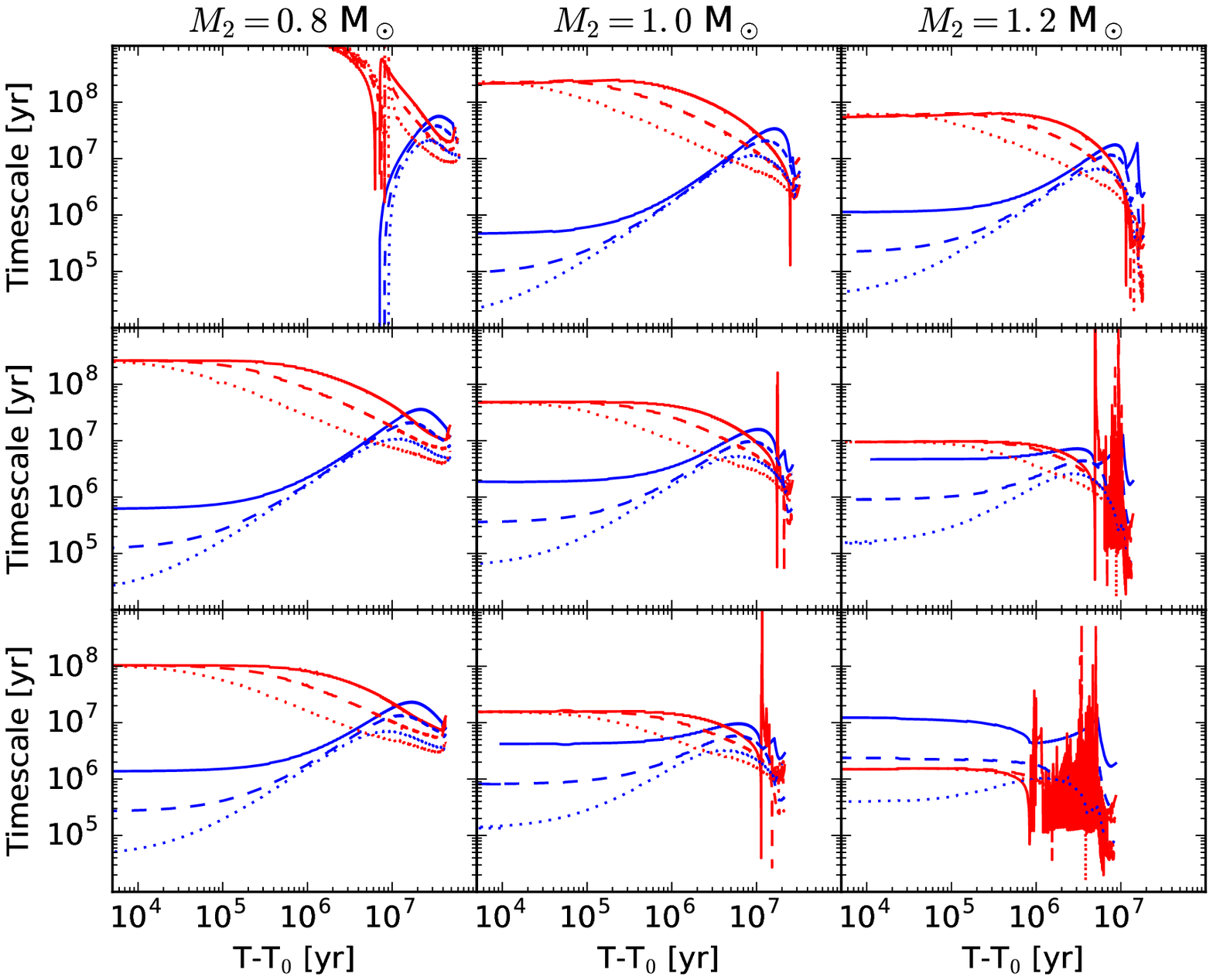}
\caption{
The evolution of $\tau_{\rm MB}$ (blue lines) and $\tau_{\rm syn}$ (red lines).
Here the input values of $M_2$ and $T_0$ and the line styles in each panel are
the same as in Fig.~\ref{f4}.
\label{f5}}
\end{figure}

To examine the competition between magnetic braking and tidal interaction in more detail,
we compare the synchronization time-scale $\tau_{\rm syn}$ and the time-scale $\tau_{\rm MB}$ of angular momentum loss due to magnetic braking, defined as
\begin{equation}
\tau_{\rm syn}=\frac{\Omega_2}{|\Omega_2-\Omega_{\rm orb}|}\tau_{\rm tid}
\end{equation}
and
\begin{equation}
\tau_{\rm MB}=\left|\frac{J_2}{\dot{J}_{\rm 2,MB}}\right|,
\end{equation}
respectively. We plot the evolution of $\tau_{\rm syn}$ and $\tau_{\rm MB}$ in Fig.~\ref{f5} with the red and blue lines, respectively. Here the input values of $M_2$ and $T_0$ and the line styles in each panel are the same as in Fig.~\ref{f4}. In most cases $\tau_{\rm syn }$ is much longer than $\tau_{\rm MB}$ initially. Then, $\tau_{\rm syn}$ starts to decrease while $\tau_{\rm MB}$ starts to increase. The break in the evolution of $\tau_{\rm MB}$ in the case of $(M_2,\,T_0)=(0.8\,M_{\odot},\,0.5\times10^7\,{\rm yr})$ is also related to the absence of the radiative core. For $(M_2,\,T_0)=(1.2\,M_{\odot},\,1.5\times10^7\,{\rm yr})$ with $f_{\rm B}=3$ and 10, $\tau_{\rm syn}$ is shorter than $\tau_{\rm MB}$ initially. In the late stage of the evolution, these two time-scales tend to be close to each other\footnote{In some cases $\tau_{\rm syn}$ changes abruptly, which is caused by numerical instability when $\Omega_{\rm 2}\rightarrow \Omega_{\rm orb}$.}.

\begin{figure}
\includegraphics[width=\columnwidth]{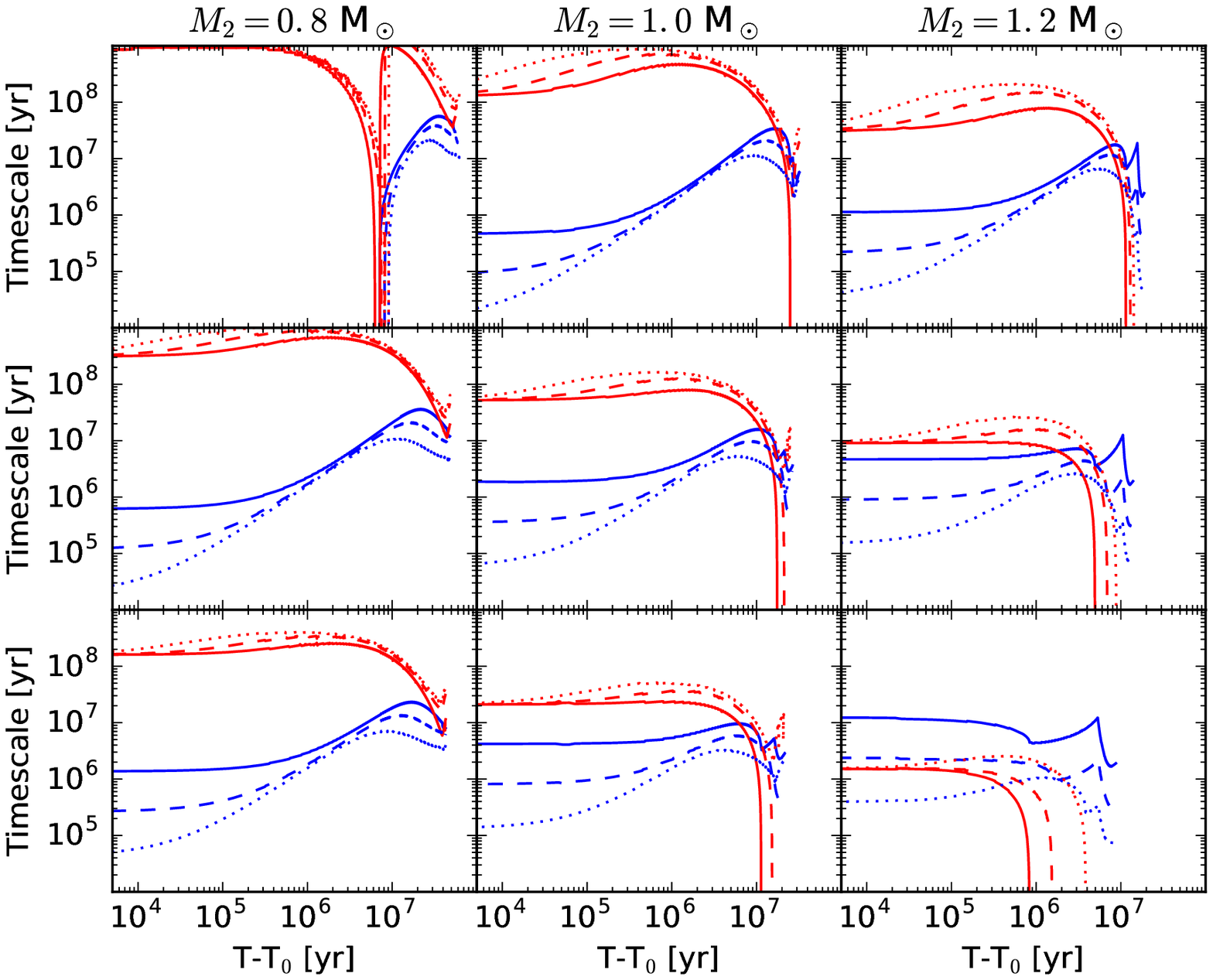}
\caption{The evolution of $\tau_{\rm tid}$ (blue lines) and $\tau_{\rm MB}$ (red lines). Here the input values of $M_2$ and $T_0$ and the line styles in each panel are the same as in Fig.~\ref{f4}.}
\label{f6}
\end{figure}

\begin{figure}
\includegraphics[width=\columnwidth]{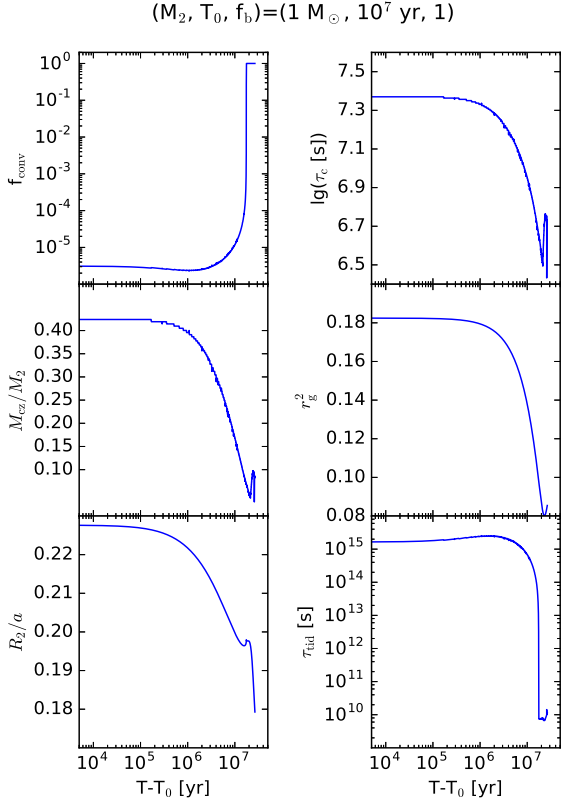}
\caption{The evolution of $f_{\rm conv}$, $\tau_{\rm c}$, $M_{\rm cz}/M_2$, $r_{\rm g}^2$, $R_2/a$, and $\tau_{\rm tid}$ in the case of $(M_2,\,T_0,\,f_{\rm B})=(1\,M_{\odot},\,1\times10^7\,{\rm yr},\,1)$.}
\label{f7}
\end{figure}

We demonstrate the evolution of $\tau_{\rm tid}$ and $\tau_{\rm MB}$ in Fig. \ref{f6} with the red and blue lines, respectively. The input values of $M_2$ and $T_0$ and the line styles in each panel are the same as in Fig.~\ref{f4}. In the early state of the evolution, $\tau_{\rm tid}$ is generally longer than $\tau_{\rm MB}$, except in the case of $(M_2,\,T_0)=(1.2\,M_{\odot},\,1.5\times10^7\,{\rm yr})$. In the late stage of the evolution, $\tau_{\rm tid}$ evolves in two trends: some evolutionary tracks end up with $\tau_{\rm tid}\sim 10^7-10^8\,{\rm yr}$, while others decrease significantly, which is mainly caused by the variation in $f_{\rm conv}$. Taking the case of $(M_2,\,T_0,\,f_{\rm B})=(1\,M_{\odot},\,1\times10^7\,{\rm yr},\,1)$ for example, we present the evolution of $f_{\rm conv}$, $\tau_{\rm c}$, $M_{\rm cz}/M_2$, $r_{\rm g}^2$, $R_2/a$, and $\tau_{\rm tid}$ in Fig. \ref{f7}. We can see that $M_{\rm cz}/M_2$, $r_{\rm g}^2$, and $R_2/a$ vary within one order of magnitude, and $\tau_{\rm c}$ decreases from $10^{7.4}\,{\rm s}$ to $10^{6.5}\,{\rm s}$. Remarkably, $f_{\rm conv}$ increases by about five orders of magnitude. Before synchronization, $f_{\rm conv}$ evolves slowly and keeps a small value, and, when $P_{\rm spin}$ approaches $P_{\rm orb}$, $f_{\rm conv}$ increases significantly. Consequently, $\tau_{\rm tid}$ decreases from $ 10^{15}\,{\rm s}$ $(\sim 10^7\,{\rm yr})$ to $10^{10}\,{\rm s}$ $(\sim 10^2\,{\rm yr})$. Therefore, the tracks associated with strong magnetic field tend to end up with long $\tau_{\rm tid}$, and  the tracks associated with weak magnetic field tend to end up with short $\tau_{\rm tid}$. Here we highlight the effect of $f_{\rm conv}$ that if the pre-MS star is not synchronized, $\tau_{\rm tid}$ is much longer than for a synchronized star even if the difference between $P_{\rm spin}$ and $P_{\rm orb}$ is small.

In summary, the tidal interaction depends on the turnover time in the pre-MS star and the difference between $P_{\rm spin}$ and $P_{\rm orb}$. Also, in pre-MS/BH binaries, tidal torque is not always able to synchronize the pre-MS star during the pre-MS phase. Therefore, under these circumstances, magnetic braking is not able to efficiently extract the orbital angular momentum.

\subsection{Mass transfer in pre-MS/BH binaries}
\begin{figure}
\includegraphics[width=\columnwidth]{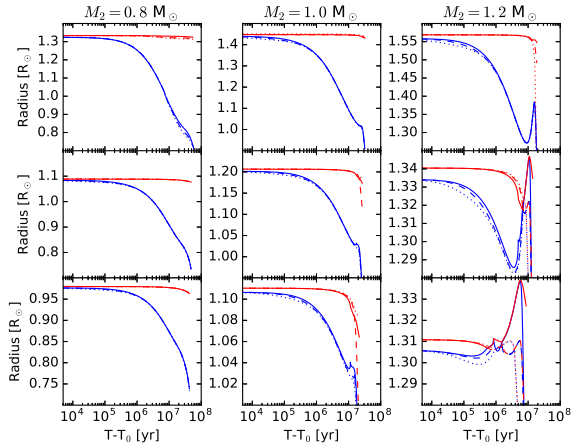}
\caption{
The evolution of $R_2$ (blue lines) and $R_{\rm L,2}$ (red lines).
Here the input values of $M_2$ and $T_0$ and the line styles in each panel are
the same as in Fig.~\ref{f4}.
\label{f8}}
\end{figure}

In this subsection, we focus on the occurrence of mass transfer in pre-MS/BH binaries and the formation of pre-MS/BHLMXB. We compare the evolution of $R_2$ and $R_{\rm L,2}$ in Fig. \ref{f8} to examine whether or not mass transfer can be triggered with the given parameters.  In Fig.~\ref{f8}, the blue and red lines represent the evolution of $R_2$ and $R_{\rm L,2}$ respectively, and the input parameters and the line styles in each panel are the same as in  Fig.~\ref{f4}. Note that in the beginning of the evolution, the pre-MS star is assumed to just fill its Roche-lobe. However, In the cases of $M_2=0.8~M_\odot$ and $1.0~M_\odot$, $R_2$ decreases more rapidly than $R_{\rm L,2}$, causing the star to be detached from its Roche-lobe. Because there is not any mass transfer, these binaries cannot evolve  into BHLMXBs. On the other hand, when $M_2=1.2~M_\odot$, the decrease in $R_2$ is followed by an increase, which indicates that after the Hayashi-line evolution, the star starts to expand. This makes $R_{\rm L,2}$ to match $R_2$ and maintain mass transfer in some cases. Generally $R_{\rm L,2}$ decreases with time because of angular momentum loss due to magnetic braking. However, in the case of $(M_2,\,T_0)=(1.2~M_\odot, 1\times 10^7\,{\rm yr})$ and $(1.2~M_\odot, 1.5\times 10^7\,{\rm yr})$, $R_{\rm L,2}$ experiences a temporary expansion phase, which is caused by rapid mass transfer between the pre-MS star and the BH.

\begin{figure}
\includegraphics[width=\columnwidth]{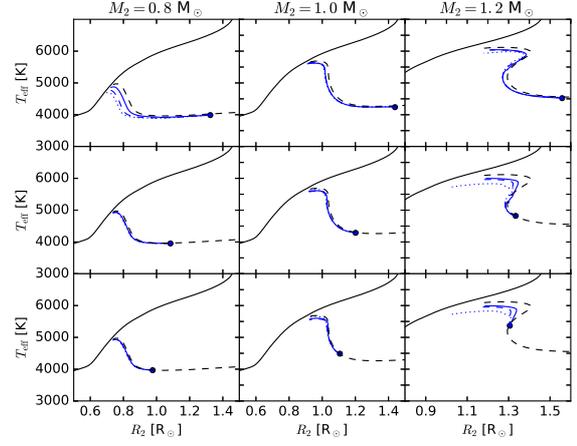}
\caption{
The evolution of the pre-MS star in binary.
In each panel of the figure, the input parameters are the same as in Fig.~\ref{f4}, the blue solid, dashed, and dotted lines represent the evolution of the pre-MS stars in binaries with $f_{\rm B}=1$, 3, and 10 respectively, the black dashed lines represent the single-star evolution,
and the black solid lines represent the MS. The blue dots denote the starting points of the binary evolution.
\label{f9}}
\end{figure}

In Fig.~\ref{f9} we compare the evolutionary tracks for the binaries in the $T_{\rm eff}-R_2$ plane. In each panel of the figure, the input parameters are the same as in Fig.~\ref{f4}, the blue solid, dashed, and dotted lines represent the evolution of the pre-MS stars in binaries with $f_{\rm B}=1$, 3, and 10 respectively, the black dashed lines represent the single-star evolution, and the black solid lines represent the MS. The blue dots denote the starting points of the binary evolution. Combining both Figs.~\ref{f8} and \ref{f9} we conclude that mass transfer can take place if the single-star evolution shows a Z-shape track, indicating the occurrence of stellar expansion. In the cases of $M_2=0.8~M_\odot$ and $1.0~M_\odot$ in which there is no mass transfer, the evolutionary tracks in binaries are still close to the single-star evolution. But for $M_2=1.2~M_\odot$, the evolutionary tracks in the mass transfer stages can significantly deviate from the single-star evolution with lower effective temperatures, which are mainly caused by mass loss of the pre-MS stars via Roche-lobe overflow.

\begin{figure}
\includegraphics[width=\columnwidth]{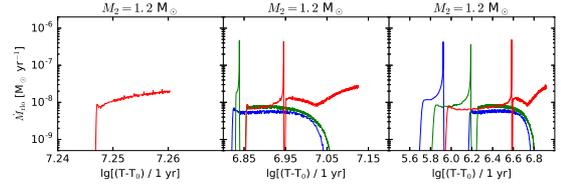}
\caption{
The evolution of the mass transfer rate $\dot{M}_{\rm rlo}$ for $M_2=1.2~M_\odot$.
In the left, middle, right panels  $T_0$ is taken to be $0.5\times 10^{7}{\rm ~yr}$, $1.0\times 10^{7}{\rm ~yr}$, and $1.5\times 10^{7}{\rm ~yr}$, respectively. The blue, green, and red lines represent $f_{\rm B}=1$, 3, and 10, respectively.
\label{f10}}
\end{figure}

Fig.~\ref{f10} shows the evolution of $\dot{M}_{\rm 2,rlo}$ for $M_2=1.2~M_\odot$. In the left, middle, right panels  $T_0$ is taken to be $0.5\times 10^{7}{\rm ~yr}$, $1.0\times 10^{7}{\rm ~yr}$, and $1.5\times 10^{7}{\rm ~yr}$, respectively. The blue, green, and red lines represent $f_{\rm B}=1$, 3, and 10, respectively. In the left panel, mass transfer occurs only when $f_{\rm B}=10$; in the middle and right panels mass transfer occurs with  $f_{\rm B}=1$, 3, and 10. The duration of the mass transfer is around $(0.5-6)\times 10^6{\rm ~yr}$. In most cases the mass transfer rates $\sim 10^{-9}-10^{-7}~M_\odot$~yr$^{-1}$, but they evolve in different trends: in the case of strong magnetic field, mass transfer rate tends to increase, while mass transfer tends to decrease in the case of weak magnetic field, which is caused by the efficiency of magnetic braking, i.e. only strong magnetic braking is able to match the rapid stellar shrink near the ZAMS. The discontinuity is caused by the rapid decrease in $\tau_{\rm c}$ when approaching the MS, where the enhanced tidal torque triggers rapid mass transfer.

\section{Discussion and Conclusions}
In this work, we examine the possible formation channel  of pre-MS/BHLMXBs proposed by \citet{Iva2006}.
In an X-ray binary system with a 7 $M_\odot$ BH and a 1 $M_\odot$ donor star, it is not unreasonable to assume the companion star was not yet on the main sequence when the BH formed, since the evolutionary time-scale of the BH progenitor (a few $10^6$ yr) is much shorter than the pre-MS phase of the low-mass donor star ($\sim 10^7$ yr). Mass transfer might be aided by orbital shrinkage caused by magnetic braking of the pre-MS star.
Following the parameterizations for the magnetic field and stellar wind of pre-MS stars suggested by \citet{Iva2006}, and taking into account the effect of the turnover time on both tidal interaction and magnetic braking, we calculate the binary evolution with MESA to examine under what conditions mass transfer can proceed in such binaries. Our results can be summarized as follows.

Our main finding is that the pre-MS stars are not always synchronized when approaching the MS. The reason is that the efficiency of tidal torque is limited by the relatively slow rotation of the low-mass star after the CE evolution and the long turnover time of the convective envelope of low-mass ($\lesssim 1.0\,M_{\odot}$) pre-MS stars of mass, which results in a too long synchronization  time in such binaries. The consequence is that magnetic braking cannot extract the orbital angular momentum efficiently, so the spin period of the pre-MS star is usually longer than the orbital period.


We further demonstrate that, the occurrence of mass transfer mainly depends on the evolution of the pre-MS star. If the stars experience expansion caused by rapid nuclear burning after the Hayashi-line evolution, then Roche-lobe overflow can take place. Such stars have masses $\gtrsim 1.2\,M_{\odot}$ and Z-shape evolutionary tracks in the $T_{\rm eff}-R$ diagram. The duration of the mass transfer is a few Myr, implying that they are short-lived X-ray sources. In addition, wind mass loss and mass transfer via Roche-lobe overflow decrease the stellar mass, so the effective temperatures are lower than in the case of single-star evolution.

\begin{figure}
\includegraphics[width=\columnwidth]{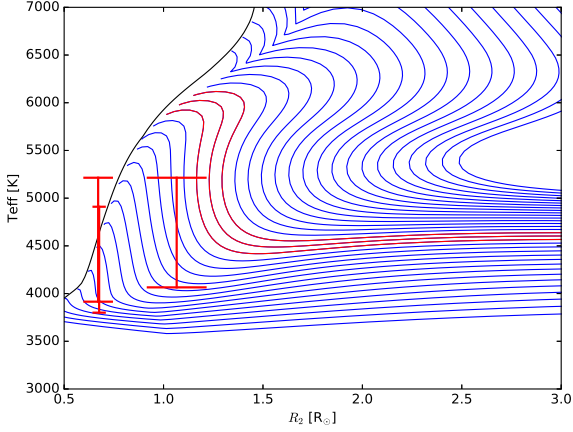}
\caption{
Single-star evolutionary tracks of the pre-MS star.
The black solid line represents the MS, and the blue and red lines represent the evolutionary tracks of pre-MS stars with mass in the range of $0.4-2~M_\odot$. We highlight the evolutionary tracks of $1.1-1.2~M_\odot$ stars with the red lines,
marking the transition region from the  Z-shape to non-Z-shape tracks.
The red horizontal and vertical bars represent the observed donor star radii and temperatures of the three systems with an observed lithium overabundance.
\label{f11}}
\end{figure}

Can the mass transferring pre-MS/BH binaries explain the BHLMXBs with Li-overabundances? We plot the single-star evolutionary tracks of pre-MS stars with different masses in the $T_{\rm eff}-R_2$ plane in Fig.~\ref{f11}. In the this figure, the black solid line represents the MS, and the blue and red lines represent the evolutionary tracks of pre-MS stars with mass in the range of $0.4-2~M_\odot$. We highlight the evolutionary tracks of $1.1-1.2~M_\odot$ stars with the red lines, marking the transition region from the  Z-shape to non-Z-shape tracks. One can see that the Z-shape feature becomes more prominent with increasing mass. The three red vertical bars represent the three BHLMXBs with  Li-overabundances (GS 2000+25, A0620-00, and Nova Muscae 1991;
data from \citet{Fra2015}\footnote{The radii of the donor stars are taken to be their Roche-lobe radii, and the effective temperatures are inferred from the spectral types of the donor stars.}).
They are all located below the red lines. This means that the pre-MS stars do not have opportunities to overflow their Roche-lobes. Therefore, these BHLMXBs cannot be explained as pre-MS/BH X-ray binaries.

\begin{figure}
\includegraphics[width=\columnwidth]{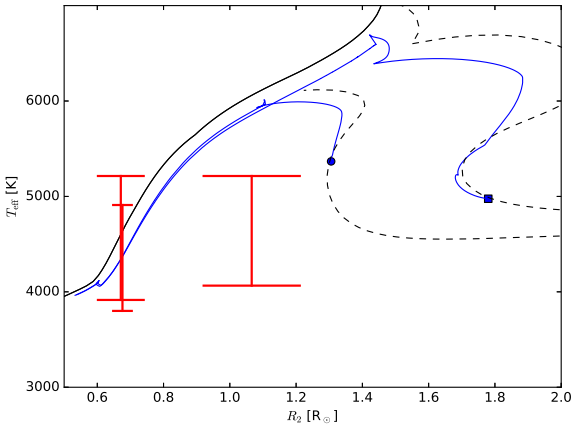}
\caption{
Evolution of pre-MS/BHLMXBs on the $T_{\rm eff}-R_2$ plane in two cases: Case 1 with $(M_{\rm BH},\,M_2,\,T_{0},\,f_{\rm B})=(7 M_{\odot},\,1.2M_{\odot},\,1.5 \times10^7{\rm ~yr},\,1)$, and Case 2 with $(M_{\rm BH},\,M_2,\,T_{0},\,f_{\rm B})=(7M_{\odot},\,1.5M_{\odot},\,0.5 \times10^7{\rm ~yr},\,1)$.
Here the black solid line represents the MS, the black dashed lines and blue solid lines represent
single-star evolution and binary evolution, respectively.
The blue dot and square denote the starting points for Cases 1 and 2, respectively.
The red horizontal and vertical bars represent the observed donor star radii and temperatures of the three systems with an observed lithium overabundance.
\label{f12}}
\end{figure}

Next, what will the mass transferring pre-MS/BH binaries evolve into? We release the termination condition $(L_{\rm nuc}/L>0.9)$ of our calculation and let the evolution of the pre-MS/BHLMXBs go on. In Fig.~\ref{f12} we show the calculated evolution on the $T_{\rm eff}-R_2$ plane in two cases: Case 1 with $(M_{\rm BH},\,M_2,\,T_{0},\,f_{\rm B})=(7 M_{\odot},\,1.2M_{\odot},\,1.5 \times10^7{\rm ~yr},\,1)$, and Case 2 with $(M_{\rm BH},\,M_2,\,T_{0},\,f_{\rm B})=(7M_{\odot},\,1.5M_{\odot},\,0.5 \times10^7{\rm ~yr},\,1)$. Here the black solid line represents the MS, the black dashed lines and blue solid lines represent single-star evolution and binary evolution, respectively. The blue dot and square denote the starting points for Cases 1 and 2, respectively. The red vertical bars represent the BHLMXBs with Li-overabundances. Mass transfer proceeds in both cases. After the pre-MS stage, the stars start to evolve along the MS with slightly lower $T_{\rm eff}$. The evolutionary time-scale of a low-mass MS star is much longer than the time-scale of orbital evolution. So the MS star has enough time to adjust its structure in response to the mass loss via Roche-lobe overflow. Therefore, we can expect the evolution will lead to the formation of an X-ray binary consisting of a low-mass MS star and an accreting BH in a short period orbit.

\section*{Acknowledgements}
This work was supported by the National Key Research and Development Program of China  (2016YFA0400803), and the Natural Science
Foundation of China (NSFC) under grant numbers 11773015 and 11333004, and Project U1838201 supported by NSFC and CAS.

\newpage

\bsp
\label{lastpage}
\end{document}